\documentclass[tightenlines,a4paper,fleqn,altaffilletter,twoside,twocolumn,
               floatfix,byrevtex]{revtex4}

\usepackage[english]{babel}
\usepackage{amsmath,amssymb}
\usepackage{graphicx}
\usepackage{hyperref}
\usepackage{url}

\newcommand{\sdm}{\ensuremath{\Delta m_{21}^2}}

\newcommand{\floor}[1]{{\lfloor #1 \rfloor}}

\begin{document}

% =============================================================================
\title{Self-Calibration of Neutrino Detectors using characteristic Backgrounds}
\author{Joachim Kopp}    \email[Email: ]{Joachim.Kopp@mpi-hd.mpg.de}
\author{Manfred Lindner} \email[Email: ]{Manfred.Lindner@mpi-hd.mpg.de}
\author{Alexander Merle} \email[Email: ]{Alexander.Merle@mpi-hd.mpg.de}
\affiliation{Max--Planck--Institut f\"ur Kernphysik\\
             Postfach 10 39 80\\
             D--69029 Heidelberg\\
             Germany}
% =============================================================================

\begin{abstract}
We introduce the possibility to use characteristic natural neutrino backgrounds, such as
Geo-neutrinos ($\bar{\nu}_e$) or solar neutrinos ($\nu_e$), with known spectral shape for the
energy calibration of future neutrino detectors, e.g. Large Liquid Scintillator Detectors.
This ``CalEffect'' could be used without the need to apply any modifications to the experiment in
all situations where one has a suitable background
with sufficient statistics. After deriving the effect analytically using $\chi^2$ statistics, we
show that it is only tiny for reactor neutrino experiments, but can be applicable in other situations.
As an example, we present its impact on the identification of the wiggles in the power
spectrum of supernova neutrinos caused by Earth matter effects. The Self-Calibration Effect could be
used for cross checking other calibration methods and to resolve systematical effects in the primary
neutrino interaction processes, in particular in the low energy cross sections.
\end{abstract}

% -----------------------------------------------------------------------------
\maketitle
% -----------------------------------------------------------------------------

% ==============================================================================
\section{Introduction}
% ==============================================================================
\vspace{-0.3cm}

As neutrino physics is entering the stage of precision measurements, future
detectors will have to face new challenges in terms of background reduction, control
of systematical uncertainties, and calibration of the counting efficiencies
and the energy reconstruction. The most widely used techniques for performing the
energy calibration are radioactive sources with known properties, artificially
accelerated lepton beams, or indirect methods based on the interaction products of
neutrinos (see e.g. Refs.~\cite{Alimonti:2000xc,Eguchi:2002dm,Nakahata:1998pz}). All of these
methods have in common that they are only sensitive to charged
particles, i.e.\ possible uncertainties associated with the primary neutrino interaction
that produces these particles are not taken into account. Therefore it is
desirable to calibrate the detector directly with a neutrino beam. Such a measurement
has been performed by GALLEX using a Cr-51 neutrino source, but it was very involved
and required a special detector design~\cite{Hampel:1997fc}. The SAGE experiment has also been calibrated with neutrinos, from Cr-51~\cite{Abdurashitov:1998ne} and Ar-37~\cite{Abdurashitov:2005tb}.

In this letter, we propose to use natural neutrino sources with known characteristic spectra, in
particular 
Geo-neutrinos and solar neutrinos as calibration sources for future low-energy neutrino
detectors. These types of neutrinos are always present, and, in a several-tens-of-kilotons detector,
they induce at least a few thousands of events per year. The energy spectra for
both, Geo-neutrinos ($\bar{\nu}_e$) and solar neutrinos ($\nu_e$), are known very precisely since
their production
processes are very well understood. They exhibit characteristic steps at the cutoff energies of
various processes, which are easy to locate in the final event spectrum even if the total rate is
not known.

Of course, the energy calibration with Geo-neutrinos or solar neutrinos will not eliminate
the need for other calibration methods, in particular because it is insensitive to
spatial or temporal variations of the detector performance and can be used only
in the very low energy region, but it is nevertheless interesting since it constitutes
a \emph{self-calibration} of the detector and provides an independent cross check of
other methods without requiring any modifications of the detector design.

In Sec.~\ref{sec:BG-Neutrinos}, we present the small effect of the detector self-calibration on
reactor neutrino experiments and explain the effect analytically. Then, in
Sec.~\ref{sec:SN}, we will show how the Self-Calibration Effect can improve the
accuracy of supernova neutrino measurements, and we finally conclude in Sec.~\ref{sec:conclusions}.

\vspace{-0.5cm}
% ==============================================================================
\section{Background Neutrinos as Calibration Sources}
% ==============================================================================
\label{sec:BG-Neutrinos}
\vspace{-0.3cm}

A future Large Liquid Scintillator Detector (LLSD) will have very good statistics even
for background neutrino sources such as distant nuclear reactors and Geo-neutrinos. This
opens the interesting possibility to use these neutrinos as calibration
sources for the energy reconstruction, which leads to the seemingly paradoxial situation
that a measurement with backgrounds can yield more precise results than a
measurement without backgrounds.

\begin{figure*}[ht]
  \begin{center}
    \includegraphics{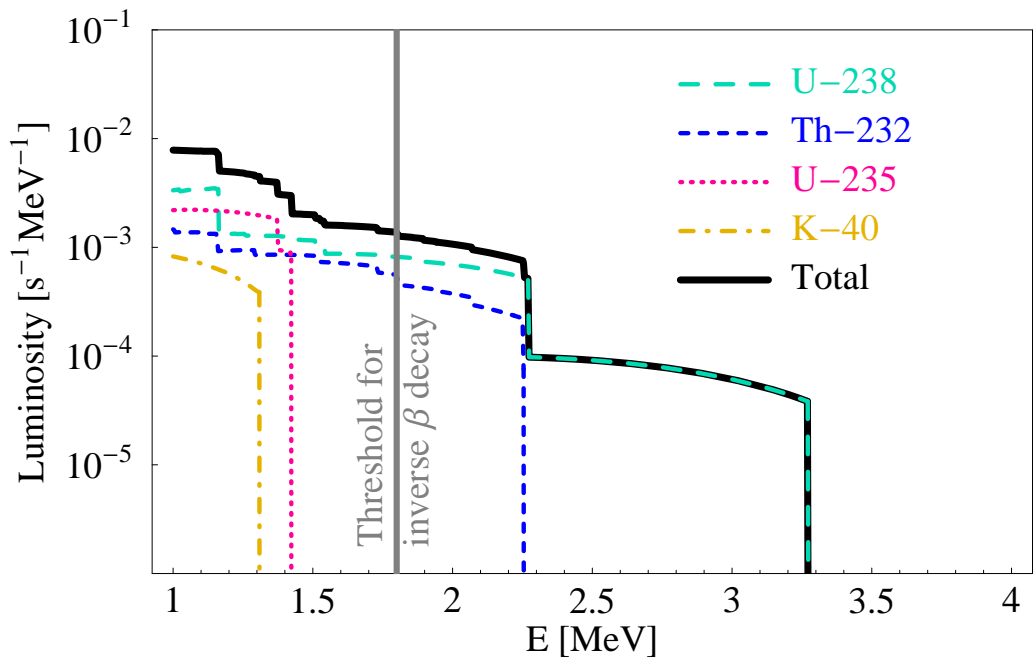}
  \end{center}
  \caption{The energy spectrum of Geo-neutrinos \cite{Geothesis,GeoHP}. Note that only
     the decay chains of uranium-238 and thorium-232 produce neutrinos that are above
     the threshold of inverse $\beta$~decay at 1.8~MeV.}
  \label{fig:geo-spectrum}
\end{figure*}

Geo-neutrinos are particularly well suited for this self-calibration because their
spectrum, which is given by the uranium and thorium decay chains, has characteristic
steps at very well known energies that can easily be located in the data samples
(see Fig.~\ref{fig:geo-spectrum}). An analogous self-calibration for $\nu_e$'s may be possible using
solar neutrinos since the spectrum of Be-8 and $pep$ is also very characteristic.

Of course, background neutrinos are insensitive to short-term variations of
the detector properties such as temperature fluctuations, so they will not
eliminate the need for other calibration methods. But they are the only tool
to resolve systematical effects in the primary neutrino interaction process
such as uncertainties in the low energy neutrino-nucleon cross sections.

In this chapter, we will first decribe in which analysis we have found the self-calibration
effect and then show that it can be explained analytically using the $\chi^2$ approach.

\vspace{-0.5cm}
% ==============================================================================
\subsection{The tiny Self-Calibration Effect in Reactor Neutrino Experiments}
% ==============================================================================
\label{sec:Reactor-Neutrinos}
\vspace{-0.3cm}

Already in~\cite{Kopp:2006mw} we have discussed the physics potential of reactor experiments
with a LLSD such as the proposed $45$~kt LENA detector~\cite{Undagoitia:1,Oberauer:2005kw}.
Using the GLoBES software~\cite{globes,Huber:2007ji}, we have investigated the accuracy of such an
experiment to measurements of the reactor angle $\theta_{13}$ as well as on the solar oscillation
parameters, $\theta_{12}$ and $\sdm$, using mobile nuclear reactors amongst others. As backgrounds, we
have taken into account the 20 closest reactors to the possible LENA site Pyh\"asalmi in Finland as
well as Geo-neutrinos coming from uranium and from thorium (the Geo-neutrinos from potassium are not
included in the analysis, since their energies are below the threshold for inverse beta decay, which
is used as detection reaction for the reactor-$\bar{\nu}_e$'s in a LLSD, see Fig.~\ref{fig:geo-spectrum}). For $\theta_{13}$, these
backgrounds turn out not to be very important: the optimum baselines for such a measurement are very
short ($\sim 1$~km), which leads, due to the large fiducial mass of the considered detector,
to extremely high event rates even for a mobile reactor. Hence, for a measurement of the small
reactor angle, the perturbation by the two different backgrounds is so small that it is not possible
to exploit any information on the spectral shape of Geo-neutrinos.

\begin{figure*}
  \begin{center}
    \includegraphics[width=13cm]{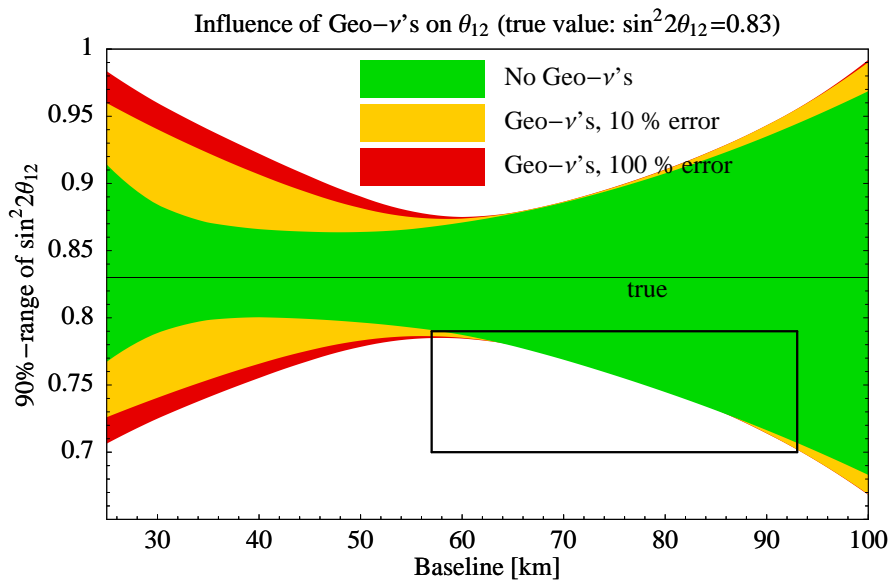}
  \end{center}
  \caption{Sensitivity of LENA to $\theta_{12}$ for the
assumed true value
as a function of the baseline where the considered reactor has a thermal power of $0.5\ {\rm
GW}_{\rm th}$. The marked part is the region
where the background self-calibration comes in.}
  \label{fig:baseline}
\end{figure*}

\begin{figure*}
  \begin{center}
    \includegraphics[width=13cm]{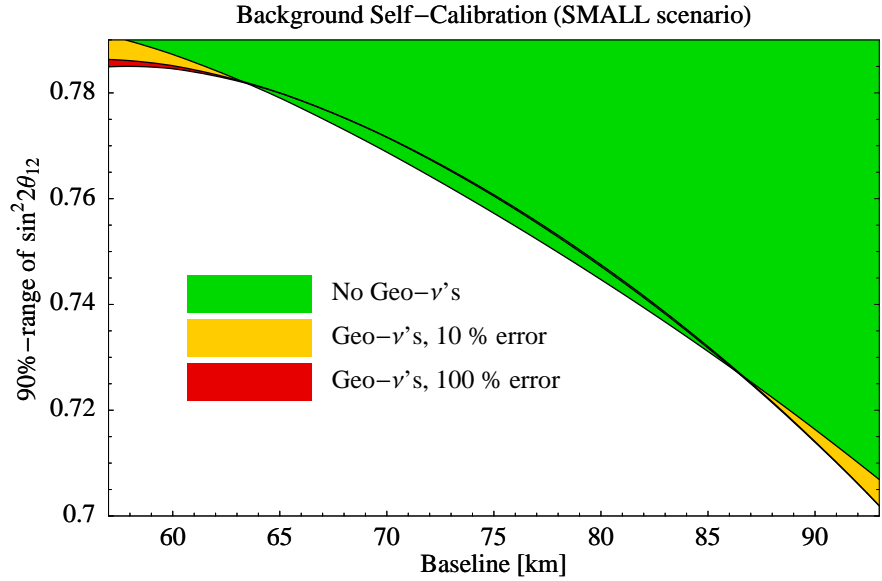}
  \end{center}
  \caption{The rectangle from Fig.~\ref{fig:baseline} drawn to a larger scale. Here, one can
clearly see the seemingly paradoxial situation that a measurement with backgrounds can yield better
results than one without.}
  \label{fig:selfcal}
\end{figure*}

However, for a measurement of $\theta_{12}$, the situation is different: considering the
\emph{SMALL} scenario (details on this analysis can be found in \cite{Kopp:2006mw}), which
corresponds to a mobile nuclear reactor with a thermal power of $0.5\ {\rm GW}_{\rm th}$ and 2~years
of data taking, one can see a small effect induced by the background self-calibration of the
experiment, which will be pointed out here. For the inclusion of Geo-neutrinos in our analysis, we
have considered three different situations:
\begin{itemize} 
  \item {\bf No Geo-neutrinos:} In this case, Geo-neutrinos are completely absent, and
    only the background from distant nuclear reactors is taken into account, which yields 850 events
per year.
  \item {\bf Geo-neutrinos with a 10\% uncertainty:} Here, Geo-neutrinos coming from uranium and 
    from thorium are taken into
    account and the uncertainty in both their fluxes is assumed to be 10\%.
    The Geo-neutrino spectra in our simulations are taken from~\cite{Geothesis,GeoHP}.
    The reactor background is the same as in the scenario without Geo-neutrinos.
  \item {\bf Geo-neutrinos with a 100\% uncertainty:} This scenario is equivalent to
    the previous one, but now the uncertainties in the two Geo-neutrino contributions
    are taken to be 100\%.
\end{itemize}
The result of our analysis for the \emph{SMALL} scenario is again plotted here in
Fig.~\ref{fig:baseline} (cf.~Fig.~3 in \cite{Kopp:2006mw}). The interesting region is the
marked rectangle. Taking a closer look it turns out to show a seemingly paradoxial situation: from
Fig.~\ref{fig:selfcal} one can see that, for certain baselines, the lower bound on the assumed true
value of $\sin^2 2\theta_{12}=0.83$ is slightly better with the background by Geo--neutrinos (and
even in both cases, with 10\% and 100\% flux uncertainty) than without. The reason is that this
background has such a characteristic form due to the well-known $Q$-values of the nuclear
production reactions of the neutrinos
that the positions of the steps in the spectrum can, in the minization of the $\chi^2$ function, be
used to reduce a systematical bias in the energy calibration essentially to zero, which may not be
possible without such a characteristic background. Hence it can indeed be the case, that, for
certain configurations, a suitable background can improve, rather than worsen, the sensitivity of
certain kinds of experiments.

The effect for reactor neutrinos is very tiny. This is because one needs to have large
background rates to be able to resolve the characteristic spectrum properly. Also the energy scale
of the background is crucial since it may not be possible to extrapolate a calibration made by MeV
neutrinos accurate enough e.g. to the GeV scale. However, the principle effect existis, as we will
demonstrate in the next section (and what we have tested using different hypothetical benchmark
scenarios), and one just has to find a scenario where this effect can be applied.

\vspace{-0.5cm}
% ------------------------------------------------------------------------------
\subsection{Analytical discussion of the Self-Calibration Effect}
% ------------------------------------------------------------------------------
\vspace{-0.3cm}

Let us now discuss the self-calibration effect analytically. We consider a simplified $\chi^2$
expression using a Gaussian approximation for the distribution of the event rates:
\begin{equation}
  \chi^2 = \sum_i^{\rm \# bins} \left[ \frac{\big(T_i(a_i, b) - N_i\big)^2}{N_i}
                        + \frac{a_i^2}{\sigma_i^2} \right].
  \label{eq:chi2ecal}
\end{equation}
Here, the $T_i$'s are the theoretically predicted event rates calculated with certain test values for the
neutrino oscillation parameters and the systematical errors, while the $N_i$'s are the rates
obtained by using the assumed ``true'' values for these parameters. The $a_i$'s are nuisance parameters as e.g.\ overall normalization errors or bin-dependent shape errors.
The explicit form of the $T_i$'s is
\begin{align}
  T_i &= (1+a_i)\tilde{N}_i (b), \label{eq:Tiecal_tot}\\
  \tilde{N}_i (b) &=(1+b)\cdot \big[ \left( N_{\floor{\delta(i)}+1}-N_{\delta(i)} \right)\cdot
    \nonumber \\
    &\cdot \left(\delta(i)-\floor{\delta(i)} \right)+N_{\floor{\delta(i)}} \big],
    \label{eq:shifted_rates}\\
  \delta(i) &= b\cdot (i+t_0+\frac{1}{2})+i. \label{eq:delta}
\end{align}
Here, $\tilde{N}_i (b)$ are the rates for wrong energy binning implied by a non-zero energy
calibration $b$ which are obtained from the correctly binned rates $N_i$ according
to Eq.~\eqref{eq:shifted_rates}, which is essentially a linear interpolation between the events in
bin
$\floor{\delta(i)}+1$ and bin $\floor{\delta(i)}$. $t_0$ is the energy threshold of
the detector, expressed in terms of the bin width, and the Gau{\ss} bracket $\floor{\cdot}$ denotes
the floor function.

The normalization uncertainties are completely contained in the bin-to-bin normalization
factors $a_i$ with errors $\sigma_i$  which contain all spectral distortion
effects, such as imperfect knowledge of the reactor signal spectrum, neutrino oscillations, etc..
The energy calibration is parametrized by $b$. Normally, one would also have to include a
penalty term for the energy calibration of the form $b^2/\sigma_b^2$ (as done for the spectral
distortions). Such a term is omitted here because we assume no external information on $b$, which
means that the energy calibration is initially completely arbitrary. We do, however, assume
$b$ to be sufficiently small to take $\floor{\delta} = i$ in Eq.~\eqref{eq:delta} (this only means that wrongly binned events are not shifted by more than one energy bin) and to
expand the whole $\chi^2$ function up to first order in $b^2$, $a_i^2$, and $b a_i$.

Combining the normalization and spectral uncertainties, $T_i$ is given by
\begin{equation}
  T_i = (1+a_i+b) \big[ (N_{i+1} - N_i) \cdot b \cdot (i + t_0 + \tfrac{1}{2}) + N_i \big].
  \label{eq:Tiecal}
\end{equation}
Neglecting terms such as $\mathcal{O}(b^3)$ and higher in $\chi^2$ means neglecting terms of order
$\mathcal{O}(b^2)$ or $\mathcal{O}(b a_i)$ in $T_i$. With this approximation, our $\chi^2$ function
becomes
\begin{align}
  \chi^2 &= \sum_i \Big( \frac{1}{N_i} \Big[ (1+a_i+b) N_i
             + b (N_{i+1} - N_i)\cdot \nonumber \\
             &\cdot(i + t_0 + \tfrac{1}{2}) - N_i \Big]^2
             + \frac{a_i^2}{\sigma_i^2} \Big).
\end{align}
This expression has to be minimized with respect to $b$:
\begin{align}
  \hspace{-1.5cm}
  \frac{\partial\chi^2}{\partial b} &= \sum_i \frac{2}{N_i} \Big[
       (a_i + b) N_i + b (N_{i+1} - N_i) (i + t_0 + \tfrac{1}{2}) \Big]\cdot \nonumber\\
       &\cdot\Big[ N_i + (N_{i+1} - N_i) (i + t_0 + \tfrac{1}{2}) \Big]= 0,  \\
  b    &= -\frac{\sum_i \frac{1}{N_i} a_i N_i \gamma_i}{\sum_i \frac{1}{N_i} \gamma_i^2},
  \label{eq:ecal}
\end{align}
where we have introduced the short hand notation
\begin{equation}
  \gamma_i = N_i + (N_{i+1} - N_i) (i + t_0 + \tfrac{1}{2}).
\end{equation}
There are two extreme cases: First, let us assume a very smooth energy spectrum,
i.e.\ $N_{i+1} - N_i \ll N_i$. Then, $\gamma_i \approx N_i$, and it follows from
Eq.~\eqref{eq:ecal} that $b$ is of the same order as $a_i$. However, if the
energy spectrum contains several large steps (as the Geo-neutrino spectrum does),
at least some of the differences $N_{i+1} - N_i$ are sizeable, so
$\gamma_i \gg N_i$. As $\gamma_i$ enters the denominator of Eq.~\eqref{eq:ecal}
quadratically, while the numerator exhibits only a linear dependence, the fit
value of $b$ will be very small in this case (close to zero), i.e.\ the energy calibration
uncertainty is essentially eliminated.

\vspace{-0.5cm}
% ------------------------------------------------------------------------------
\section{Application to Earth matter effects on supernova neutrinos}
% ------------------------------------------------------------------------------
\label{sec:SN}
\vspace{-0.3cm}

On the one hand, the self-calibration effect is a nice tool to make an
experimentalist's work more efficient. But to demonstrate that it can also help to do
precision measurements, we consider the matter effects of the Earth on supernova
neutrinos as an example using the same LLSD as in Sec.~\ref{sec:BG-Neutrinos}.

The fact that Earth matter has an effect on the spectrum of supernova neutrinos has been presented
in~\cite{Dighe:2003jg,Dighe:2003vm}. A supernova core is essentially a neutrino blackbody source,
so that the spectrum of $\overline{\nu}_e$'s produced in the supernova is thermal.
However, it is important to keep in mind that the neutrinos that are produced as $\overline{\nu}_e$
reach us as matter eigenstates $\overline{\nu}_1$, which means that the oscillations that
appear inside the Earth are $\overline{\nu}_1$--$\overline{\nu}_2$ oscillations~\cite{Dighe:1999bi}.
This has been taken into account in our simulations by modifying the source code of the
{\sf GLoBES} software accordingly.

The effect of the oscillations in the Earth on the neutrino energy spectrum is then, that one can see
wiggles on the otherwise smooth spectrum. This can be shown more clearly by going at
first to inverse energy units and then taking the squared modulus of the Fourier transform of
the resulting spectrum, which gives the power spectrum, that points out the different oscillation
modes. In the case of Earth matter effects, this results in one or more peaks, depending on whether
the neutrinos traverse the core of the Earth or only the mantle. To extract these peaks from
the experimental data, it is important to precisely know their positions. This obviously requires
an excellent energy calibration~\cite{Dighe:2003vm}.

As example for our calculations, we have taken the ``accretion-phase model I'' from~\cite{Keil:2002in}.
This gives the value $\alpha=4.4$ for the flux parameter as well as
$\Phi(\overline{\nu}_e)/\Phi(\overline{\nu}_x)=0.8$ (for details, see~\cite{Dighe:2003vm}), where $x$ stands for all
other neutrino flavours except $\overline{\nu}_e$. For modelling the Earth matter, we have taken
two different scenarios: a constant density equal to the average mantle density of $4.6\
\textrm{g}/\textrm{cm}^3$ according to the PREM profile~\cite{Dziewonski:1981xy}
as well as a 3~layer approximation with two times the mantle density, but a
different core density of $11.8\ \textrm{g}/\textrm{cm}^3$. In both scenarios, the neutrinos travel
through the whole Earth, which corresponds to a baseline of $12742$ km. Note that our normalization
is made for approximately 2000 events in the detector, which is a good example value according
to~\cite{Dighe:2003jg}.

\begin{figure*}[ht]
  \begin{center}
    \includegraphics[width=8.8cm]{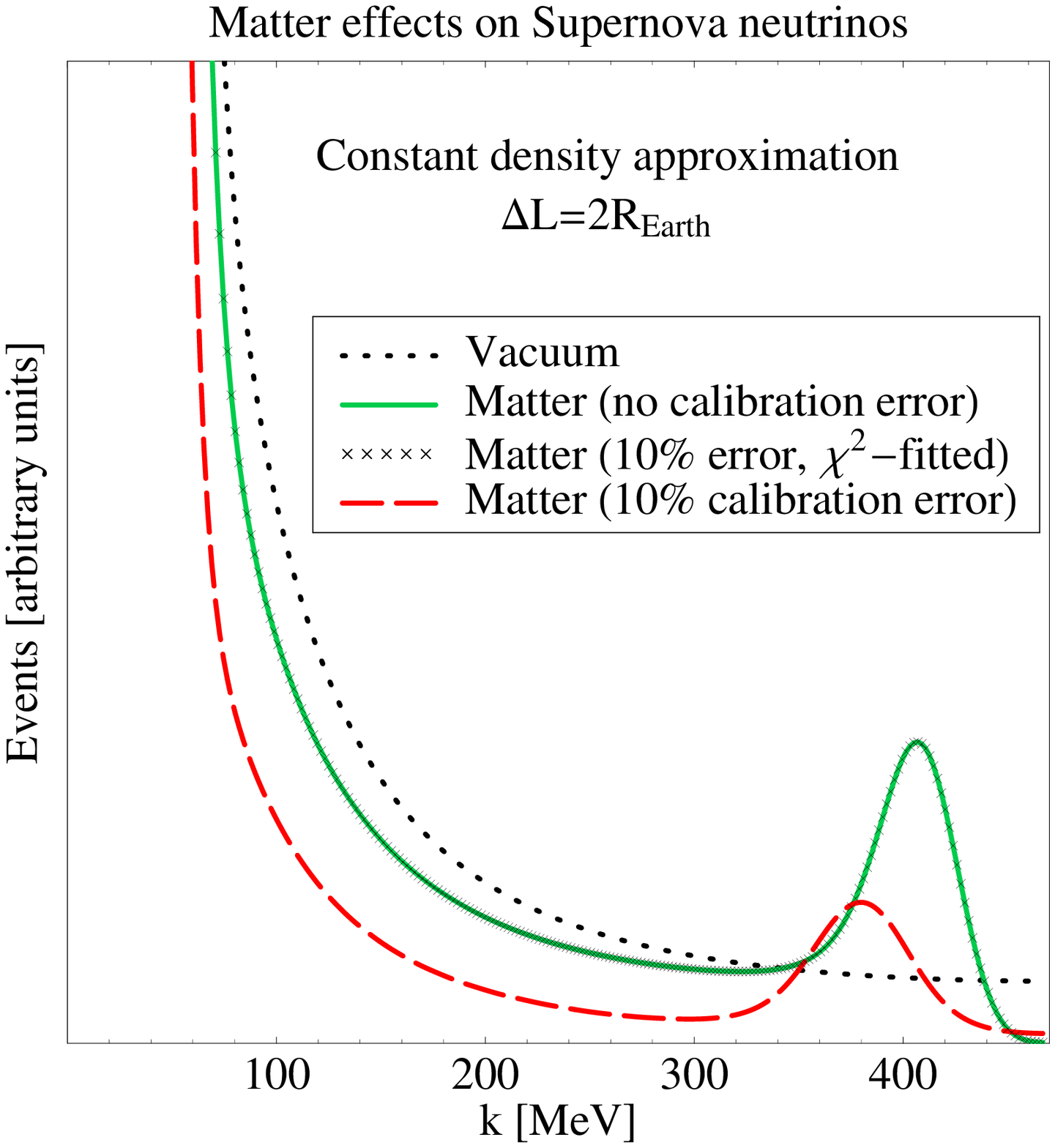}
    \includegraphics[width=8.8cm]{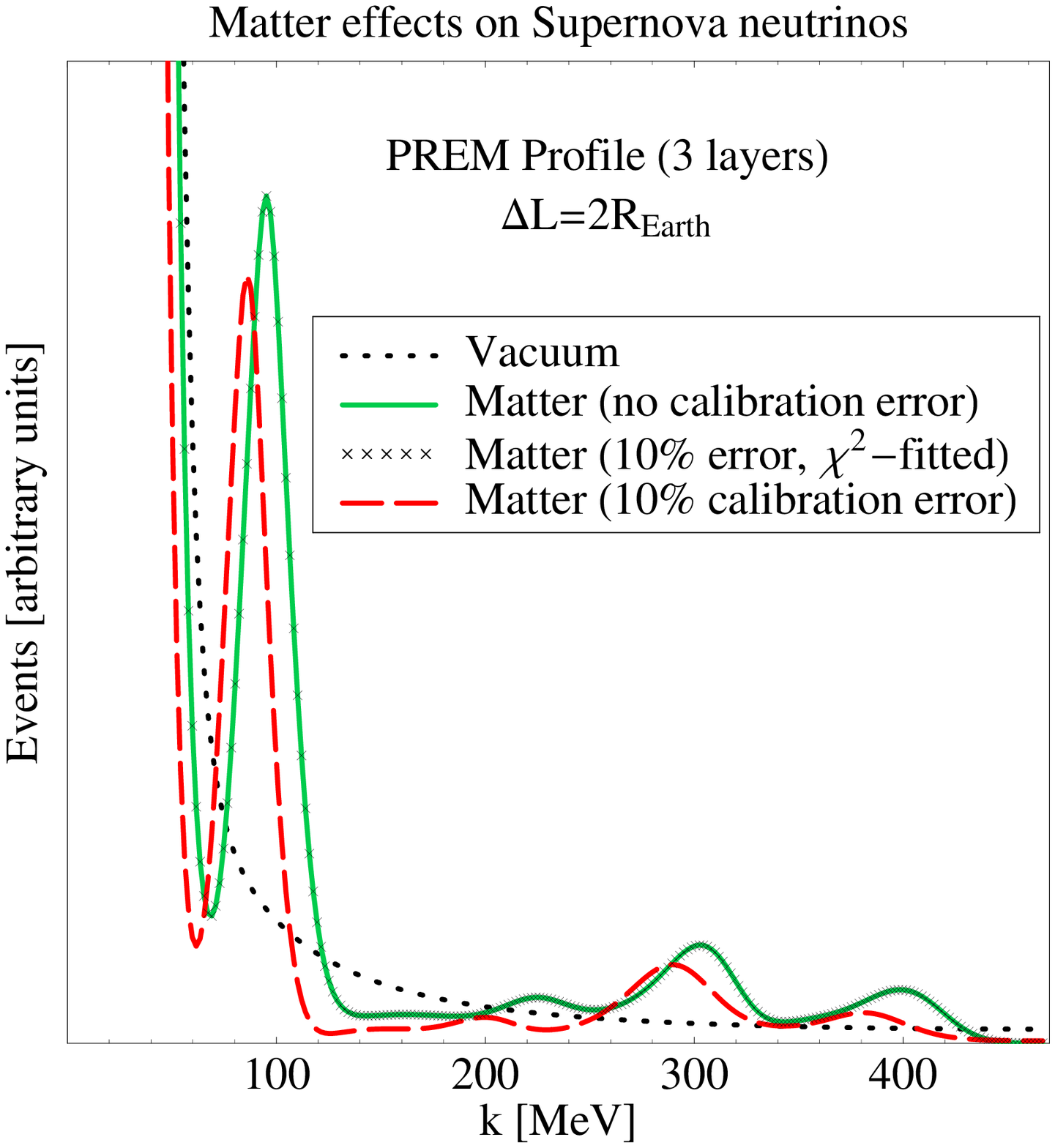}
  \end{center}
  \caption{Matter effects on the power spectrum of supernova neutrinos for a constant density
profile and a 3-layer approximation. The neutrinos always propagate through the whole diameter of
the Earth. Considered are the cases of propagation through vacuum and through matter, while the
latter is subdivided in the case with perfect energy calibration and the one with a 10\%
calibration error. Furthermore, for the wrong calibration, we have in one case performed a
$\chi^2$ analysis and shifted the events with the fittet value of the energy calibration. Due to the
self-calibration effect, this analysis has again given the correct adjustment of the energy, which
causes the corrected data points to lie exactly on the line of perfect energy calibration.}
  \label{fig:SN}
\end{figure*}

This has been simulated with the {\sf GLoBES} software package for different cases (cf.
Fig.~\ref{fig:SN}). The first case is propagation in vacuum (black-dotted line). This
does of course not have any effect on the power spectrum of neutrinos reaching the Earth from the
supernova, simply because it is completey
equivalent to a $12742$ km longer propagation in space, which clearly makes no difference for a
supernova whose expected distance from the Earth is several kpc. For propagation through matter, one
can see - as expected - one or even more peaks, depending on the number of layers of constant density in
the considered Earth model. The red-dashed curve with a wrong energy calibration (10\% error) is
clearly separated from the one with perfect calibration (green-solid). Of course, 10\% is much
worse than this error would be in reality, but here just the Self-Calibration Effect is to be
demonstrated which is more illustrative for a larger error. The crosses are data points coming from
a simulation that has also been started with an initial calibration error of 10\%, but in that case,
we have fitted the energy calibration $b$ to the Geo-neutrino background. This can be done since the
SN neutrinos can be easily separated from all other
events simply because of the narrow time window for neutrinos coming from a supernova. This $\chi^2$
analysis then pulls the value of the energy calibration $b$ down to zero because of the
self-calibration
effect. Then, shifting the energy of the events by the best-fit value of the calibration gives no
difference to the case of perfect energy calibration, which can be seen for both scenarios in
Fig.~\ref{fig:SN}.

Note that in this calculation, we have assumed a perfect extrapolation of the energy calibration to
higher energies. Geo-neutrinos have energies up to about $3.3$~MeV (cf.
Fig.~\ref{fig:geo-spectrum}), while supernova neutrinos can have energies of several tens of MeV.
Of course, in reality, this extrapolation cannot work perfectly, but may be possible to a good
enough precision.

This shows that the Self-Calbration Effect can be a nice tool for measuring Earth matter effects
on neutrinos produced in
supernovae. However, this is just one example, but of course, the effect applies
to all situations, where an accurate energy calibration is required and a background source with
characteristic spectrum and good enough statistics is available. Possible applications of the
solar neutrino spectrum should also be considered.

\vspace{-0.5cm}
% =============================================================================
\section{Conclusions}
% =============================================================================
\label{sec:conclusions}
\vspace{-0.3cm}

We have presented the interesting possibility for a Large Liquid Scintillator Detector in future
neutrino experiments to use natural backgrounds with a characteristic spectral shape for the energy
calibration. Due to their well-known spectrum, this could be Geo-neutrinos as source of
$\bar{\nu}_e$'s or solar neutrinos as source of $\nu_e$'s. We have demonstrated that this effect
is only tiny for reactor neutrino experiments due to their high statistics and known spectral
shape. However, we have also shown that the effect can be derived analytically and that it could be
a
nice tool in suitable scenarios, e.g.~supernovae, where the correct energy calibration is
crucial for detecting Earth matter effects on the neutrinos originating from such an explosion. The
effect applies to all situations with a suitable background source and high enough statistics.

\vspace{-0.5cm}
% =============================================================================
\section*{Acknowledgments}
% =============================================================================

We would like to thank K.~Hochmuth, W.~Hampel, P.~Huber, T.~Marrod\'an-Undagoitia, M.~Rolinec, M.~Wurm, and
especially
L.~Oberauer for useful discussions and information on the LENA detector as well as H.-T.~Janka,
G.~Raffelt, and
R.~Tom\`as for useful comments and information. We are grateful to S.~Enomoto for sending
us his data files on Geo-neutrino spectra in machine-readable form. This work has been supported by
the Sonderforschungsbereich TR27 ``Neutrinos and Beyond'' der Deutschen Forschungsgemeinschaft.

% =============================================================================
\bibliography {caleffect}

\begin{thebibliography}{18}
\expandafter\ifx\csname natexlab\endcsname\relax\def\natexlab#1{#1}\fi
\expandafter\ifx\csname bibnamefont\endcsname\relax
  \def\bibnamefont#1{#1}\fi
\expandafter\ifx\csname bibfnamefont\endcsname\relax
  \def\bibfnamefont#1{#1}\fi
\expandafter\ifx\csname citenamefont\endcsname\relax
  \def\citenamefont#1{#1}\fi
\expandafter\ifx\csname url\endcsname\relax
  \def\url#1{\texttt{#1}}\fi
\expandafter\ifx\csname urlprefix\endcsname\relax\def\urlprefix{URL }\fi
\providecommand{\bibinfo}[2]{#2}
\providecommand{\eprint}[2][]{\url{#2}}

\bibitem[{\citenamefont{Alimonti et~al.}(2002)}]{Alimonti:2000xc}
\bibinfo{author}{\bibfnamefont{G.}~\bibnamefont{Alimonti}} \bibnamefont{et~al.}
  (\bibinfo{collaboration}{Borexino}), \bibinfo{journal}{Astropart. Phys.}
  \textbf{\bibinfo{volume}{16}}, \bibinfo{pages}{205} (\bibinfo{year}{2002}),
  \eprint{hep-ex/0012030}.

\bibitem[{\citenamefont{Eguchi et~al.}(2003)}]{Eguchi:2002dm}
\bibinfo{author}{\bibfnamefont{K.}~\bibnamefont{Eguchi}} \bibnamefont{et~al.}
  (\bibinfo{collaboration}{KamLAND}), \bibinfo{journal}{Phys. Rev. Lett.}
  \textbf{\bibinfo{volume}{90}}, \bibinfo{pages}{021802}
  (\bibinfo{year}{2003}), \eprint{hep-ex/0212021}.

\bibitem[{\citenamefont{Nakahata et~al.}(1999)}]{Nakahata:1998pz}
\bibinfo{author}{\bibfnamefont{M.}~\bibnamefont{Nakahata}} \bibnamefont{et~al.}
  (\bibinfo{collaboration}{Super-Kamiokande}), \bibinfo{journal}{Nucl. Instrum.
  Meth.} \textbf{\bibinfo{volume}{A421}}, \bibinfo{pages}{113}
  (\bibinfo{year}{1999}), \eprint{hep-ex/9807027}.

\bibitem[{\citenamefont{Hampel et~al.}(1998)}]{Hampel:1997fc}
\bibinfo{author}{\bibfnamefont{W.}~\bibnamefont{Hampel}} \bibnamefont{et~al.}
  (\bibinfo{collaboration}{GALLEX}), \bibinfo{journal}{Phys. Lett.}
  \textbf{\bibinfo{volume}{B420}}, \bibinfo{pages}{114} (\bibinfo{year}{1998}).

\bibitem[{\citenamefont{Abdurashitov et~al.}(1999)}]{Abdurashitov:1998ne}
\bibinfo{author}{\bibfnamefont{J.~N.} \bibnamefont{Abdurashitov}}
  \bibnamefont{et~al.} (\bibinfo{collaboration}{SAGE}), \bibinfo{journal}{Phys.
  Rev.} \textbf{\bibinfo{volume}{C59}}, \bibinfo{pages}{2246}
  (\bibinfo{year}{1999}), \eprint{hep-ph/9803418}.

\bibitem[{\citenamefont{Abdurashitov et~al.}(2006)}]{Abdurashitov:2005tb}
\bibinfo{author}{\bibfnamefont{J.~N.} \bibnamefont{Abdurashitov}}
  \bibnamefont{et~al.}, \bibinfo{journal}{Phys. Rev.}
  \textbf{\bibinfo{volume}{C73}}, \bibinfo{pages}{045805}
  (\bibinfo{year}{2006}), \eprint{nucl-ex/0512041}.

\bibitem[{\citenamefont{Kopp et~al.}(2007)\citenamefont{Kopp, Lindner, Merle,
  and Rolinec}}]{Kopp:2006mw}
\bibinfo{author}{\bibfnamefont{J.~F.} \bibnamefont{Kopp}},
  \bibinfo{author}{\bibfnamefont{M.}~\bibnamefont{Lindner}},
  \bibinfo{author}{\bibfnamefont{A.}~\bibnamefont{Merle}}, \bibnamefont{and}
  \bibinfo{author}{\bibfnamefont{M.}~\bibnamefont{Rolinec}},
  \bibinfo{journal}{JHEP} \textbf{\bibinfo{volume}{01}}, \bibinfo{pages}{053}
  (\bibinfo{year}{2007}), \eprint{hep-ph/0606151}.

\bibitem[{\citenamefont{Marrodan-Undagoitia et~al.}(2005)}]{Undagoitia:1}
\bibinfo{author}{\bibfnamefont{T.}~\bibnamefont{Marrodan-Undagoitia}}
  \bibnamefont{et~al.}, \bibinfo{journal}{Phys. Rev.}
  \textbf{\bibinfo{volume}{D72}}, \bibinfo{pages}{075014}
  (\bibinfo{year}{2005}), \eprint{hep-ph/0511230}.

\bibitem[{\citenamefont{Oberauer et~al.}(2005)\citenamefont{Oberauer, von
  Feilitzsch, and Potzel}}]{Oberauer:2005kw}
\bibinfo{author}{\bibfnamefont{L.}~\bibnamefont{Oberauer}},
  \bibinfo{author}{\bibfnamefont{F.}~\bibnamefont{von Feilitzsch}},
  \bibnamefont{and} \bibinfo{author}{\bibfnamefont{W.}~\bibnamefont{Potzel}},
  \bibinfo{journal}{Nucl. Phys. Proc. Suppl.} \textbf{\bibinfo{volume}{138}},
  \bibinfo{pages}{108} (\bibinfo{year}{2005}).

\bibitem[{\citenamefont{Huber et~al.}(2005)\citenamefont{Huber, Lindner, and
  Winter}}]{globes}
\bibinfo{author}{\bibfnamefont{P.}~\bibnamefont{Huber}},
  \bibinfo{author}{\bibfnamefont{M.}~\bibnamefont{Lindner}}, \bibnamefont{and}
  \bibinfo{author}{\bibfnamefont{W.}~\bibnamefont{Winter}},
  \bibinfo{journal}{Comput. Phys. Commun.} \textbf{\bibinfo{volume}{167}},
  \bibinfo{pages}{195} (\bibinfo{year}{2005}), \eprint{hep-ph/0407333}.

\bibitem[{\citenamefont{Huber et~al.}(2007)\citenamefont{Huber, Kopp, Lindner,
  Rolinec, and Winter}}]{Huber:2007ji}
\bibinfo{author}{\bibfnamefont{P.}~\bibnamefont{Huber}},
  \bibinfo{author}{\bibfnamefont{J.}~\bibnamefont{Kopp}},
  \bibinfo{author}{\bibfnamefont{M.}~\bibnamefont{Lindner}},
  \bibinfo{author}{\bibfnamefont{M.}~\bibnamefont{Rolinec}}, \bibnamefont{and}
  \bibinfo{author}{\bibfnamefont{W.}~\bibnamefont{Winter}}
  (\bibinfo{year}{2007}), \eprint{hep-ph/0701187}.

\bibitem[{\citenamefont{Enomoto}(2005)}]{Geothesis}
\bibinfo{author}{\bibfnamefont{S.}~\bibnamefont{Enomoto}}, Ph.D. thesis
  (\bibinfo{year}{2005}).

\bibitem[{\citenamefont{Enomoto}()}]{GeoHP}
\bibinfo{author}{\bibfnamefont{S.}~\bibnamefont{Enomoto}},
  \urlprefix\nolinkurl{http://www.awa.tohoku.ac.jp/~sanshiro/geoneutrino/spectrum/}.

\bibitem[{\citenamefont{Dighe et~al.}(2003)\citenamefont{Dighe, Keil, and
  Raffelt}}]{Dighe:2003jg}
\bibinfo{author}{\bibfnamefont{A.~S.} \bibnamefont{Dighe}},
  \bibinfo{author}{\bibfnamefont{M.~T.} \bibnamefont{Keil}}, \bibnamefont{and}
  \bibinfo{author}{\bibfnamefont{G.~G.} \bibnamefont{Raffelt}},
  \bibinfo{journal}{JCAP} \textbf{\bibinfo{volume}{0306}}, \bibinfo{pages}{006}
  (\bibinfo{year}{2003}), \eprint{hep-ph/0304150}.

\bibitem[{\citenamefont{Dighe et~al.}(2004)\citenamefont{Dighe, Kachelriess,
  Raffelt, and Tomas}}]{Dighe:2003vm}
\bibinfo{author}{\bibfnamefont{A.~S.} \bibnamefont{Dighe}},
  \bibinfo{author}{\bibfnamefont{M.}~\bibnamefont{Kachelriess}},
  \bibinfo{author}{\bibfnamefont{G.~G.} \bibnamefont{Raffelt}},
  \bibnamefont{and} \bibinfo{author}{\bibfnamefont{R.}~\bibnamefont{Tomas}},
  \bibinfo{journal}{JCAP} \textbf{\bibinfo{volume}{0401}}, \bibinfo{pages}{004}
  (\bibinfo{year}{2004}), \eprint{hep-ph/0311172}.

\bibitem[{\citenamefont{Dighe and Smirnov}(2000)}]{Dighe:1999bi}
\bibinfo{author}{\bibfnamefont{A.~S.} \bibnamefont{Dighe}} \bibnamefont{and}
  \bibinfo{author}{\bibfnamefont{A.~Y.} \bibnamefont{Smirnov}},
  \bibinfo{journal}{Phys. Rev.} \textbf{\bibinfo{volume}{D62}},
  \bibinfo{pages}{033007} (\bibinfo{year}{2000}), \eprint{hep-ph/9907423}.

\bibitem[{\citenamefont{Keil et~al.}(2003)\citenamefont{Keil, Raffelt, and
  Janka}}]{Keil:2002in}
\bibinfo{author}{\bibfnamefont{M.~T.} \bibnamefont{Keil}},
  \bibinfo{author}{\bibfnamefont{G.~G.} \bibnamefont{Raffelt}},
  \bibnamefont{and} \bibinfo{author}{\bibfnamefont{H.-T.} \bibnamefont{Janka}},
  \bibinfo{journal}{Astrophys. J.} \textbf{\bibinfo{volume}{590}},
  \bibinfo{pages}{971} (\bibinfo{year}{2003}), \eprint{astro-ph/0208035}.

\bibitem[{\citenamefont{Dziewonski and Anderson}(1981)}]{Dziewonski:1981xy}
\bibinfo{author}{\bibfnamefont{A.~M.} \bibnamefont{Dziewonski}}
  \bibnamefont{and} \bibinfo{author}{\bibfnamefont{D.~L.}
  \bibnamefont{Anderson}}, \bibinfo{journal}{Phys. Earth Planet. Interiors}
  \textbf{\bibinfo{volume}{25}}, \bibinfo{pages}{297} (\bibinfo{year}{1981}).

\end{thebibliography}
% =============================================================================

\newpage

\end{document}